\documentclass[conference]{IEEEtran}

\IEEEoverridecommandlockouts                              
\usepackage{graphics}       
\usepackage{epsfig}         
\usepackage{times}          
\usepackage{amsmath}        
\usepackage{amssymb}        
\usepackage{color}
\usepackage{array}

\begin{document}

\title{Full- and Reduced-order Model of Hydraulic Cylinder for Motion Control}

\author{\IEEEauthorblockN{Michael Ruderman}
\IEEEauthorblockA{University of Agder (UiA), Post box 422,
4604-Kristiansand, Norway \\ Email: \tt\small
michael.ruderman@uia.no}
\thanks{\textcolor[rgb]{0.00,0.00,1.00}{Preprint of the manuscript accepted to IEEE 43rd Annual
Conference of Industrial Electronics Society (IECON2017)}}
}

\maketitle
\thispagestyle{empty}
\pagestyle{empty}

\begin{abstract}
This paper describes the full- and reduced-order models of an
actuated hydraulic cylinder suitable for system dynamics analysis
and motion control design. The full-order model incorporates the
valve spool dynamics with combined dead-zone and saturation
nonlinearities -- inherent for the orifice flow. It includes the
continuity equations of hydraulic circuits coupled with the
dynamics of mechanical part of cylinder drive. The resulted model
is the fifth-order and nonlinear in states. The reduced model
neglects the fast valve spool dynamics, simplifies both the
orifice and continuity equations through an aggregation, and
considers the cylinder rod velocity as output of interest. The
reduced model is second-order that facilitates studying the system
behavior and allows for direct phase plane analysis. Dynamics
properties are addressed in details, for both models, with focus
on the frequency response, system damping, and state trajectories
related to the load pressure and relative velocity.
\end{abstract}

\section{INTRODUCING REMARK}
\label{sec:1}

Hydraulic cylinders \cite{merritt1967hydraulic} are the first
choice in numerous applications which require high forces and
robust operation in outside, harsh, and hard-accessible
environments. Motion control \cite{yao2000adaptive}, force control
\cite{alleyne2000simplified}, and hybrid of both
\cite{komsta2013integral} represent steady challenging tasks for
the operation of hydraulic cylinders. Due to a complex nonlinear
behavior, their reliable modeling, identification, and control
\cite{sohl1999experiments} are crucial for subsequent
exploitation. This note addresses the full- and reduced-order
modeling of hydraulic cylinders seen from the system plant point
of view required for the analysis and control design.

\section{FULL-ORDER MODEL}
\label{sec:2}

\begin{figure}[!h]
\centering
\includegraphics[width=0.62\columnwidth]{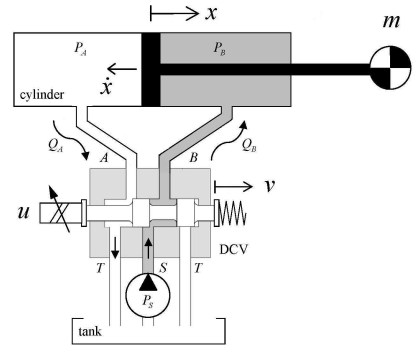}
\caption{Principal structure of hydraulic cylinder controlled by
DCV.} \label{fig:1}
\end{figure}

We first consider the full-order model of hydraulic cylinder
controlled by a directional control valve (DCV). Note that the
following modeling relies on the basics of hydraulic servo
systems, provided e.g. in \cite{merritt1967hydraulic}, and is
close to developments which can also be founded in the previously
published works on hydraulic servo systems, e.g.
\cite{sohl1999experiments,komsta2013integral}. However, whenever
differences appear we will highlight these with the corresponding
references. The DCV allows for a volumetric flow $Q_n$ of
hydraulic medium from, correspondingly to, both chambers (A and B)
of hydraulic cylinder, see Fig. \ref{fig:1}. Note that here and
later on the subscript $n=\{A, B, S, T\}$ refers to the cylinder
chambers A and B, hydraulic pressure supply (for instance pump),
and the tank correspondingly. The pressure difference between both
chambers generates the hydraulic actuation force which moves the
cylinder piston with a rod in direction of the pressure gradient.
The external mechanical loads, inertial force, and cylinder
friction counteract the induced pressure-gradient force, thus
resulting in the motion dynamics of cylinder drive.

In the standard configuration with a constant supply pressure
$P_S$, the single available control input is that of the DCV
denoted by $u$. Since the DCV spool is actuated by a proportional
solenoid, the choice of the $u$ quantity strictly depends on the
way how the solenoid, and correspondingly DCV, is low-level
controlled. Most simple case, $u$ is the coil voltage applied
through an amplifier so as to energize the electro-magnetic
circuits of the solenoid. However, the more advanced and common
case, which we also assume in the following, is when the control
input $u$ is directly proportional to the relative spool position.
That means the latter is low-level controlled by an embedded DCV
electronics, which includes an internal coil current loop and
external spool position loop. Note that the spool relative
displacement $\nu$ is provided by either a single or by a pair of
electro-magnetic solenoids. In the first case, a pre-stressed
returning spring ensures the bidirectional spool actuation, cf.
with Fig. \ref{fig:1}. In the second case, two synchronized
solenoids are arranged in antagonistic way, thus making spool
pushing and pulling fully identical.

The low-level controlled DCV has the closed-loop transfer
characteristics of the second-order system described by
\begin{equation}
G(s) = \frac{\nu(s)}{u(s)} = \frac{\omega_0^2}{s^2 + 2 \xi
\omega_0 s + \omega_0^2}. \label{eq:e1}
\end{equation}
The closed-loop parameters are the eigen-frequency $\omega_0$,
damping ratio $\xi$ and, ideal-case, unity gain which ensures the
steady-state accuracy of the spool position control loop.
Depending on the input amplitude $|u|$ and working pressure in
hydraulic circuits, the transfer characteristics $G(s)$ can be
varying in parameters and, therefore, bear some minor
uncertainties in transient behavior of the controlled spool
position response. Some typical $|u|$-dependent variations in the
nominal (measured) frequency response characteristics of the
controlled DCV are in vicinity to corner frequencies, and
therefore DCV control bandwidth. This can be found in several
manufacturer data sheets of the commercially available DCVs with
embedded electronics that implements a low-level spool position
control.

When the DCV assembly is equipped with a closed center spool,
which is one of the most common configurations aimed for reducing
the valve leakage and sensitivity to the small input signals, the
spool-controlled flow characteristics are inherently subject to
the dead-zone nonlinearity. That is a relative spool displacement
below certain magnitude, in both directions across (zero) center
position, does not lead to an orifice and therefore hydraulic
flow. Furthermore, the orifice, which is governed by the spool
displacement, is subject to saturations due to the maximally
possible opening given by the inner structure of the valve body.
Therefore, an internal flow-governing control variable $z$ can be
introduced as being related to the controlled spool position
through two static nonlinearities, dead-zone and saturation,
connected in series. Here we note that several previous works
either neglect or only partially account for the above
nonlinearities, while these can have an impact on the overall
system dynamics in view of cylinder feedback control in the
applications. So e.g. in
\cite{sohl1999experiments,alleyne2000simplified,yao2000adaptive,marton2011practical,komsta2013integral}
both nonlinearities are neglected. In
\cite{aranovskiy2013modeling} both nonlinearities are taken into
account while disregardind the second-order DCV closed-loop
dynamics and approximating the spool position response as a
first-order system. A more detailed and sophisticated model of
proportional valves with internal flow dynamics
\cite{eryilmaz2006unified} explicitly analyzed this type of
nonlinearities. However the spool travel $\nu$ has been assumed as
an available external input, therefore without considering its
controlled behavior and associated (internal) dynamics.

Combining the DCV closed-loop dynamics (\ref{eq:e1}) with the
overall static nonlinearity described by
\begin{equation}
h(\nu) = \left\{%
\begin{array}{ll}
    \alpha \, \mathrm{sign}(\nu),      & \hbox{ if } |\nu| \geq \alpha+\beta,
    \\[1mm]
    0,                              & \hbox{ if } |\nu| < \beta,
    \\[1mm]
    \nu - \beta \, \mathrm{sign}(\nu), & \hbox{ otherwise}, \\
\end{array}%
\right.      \label{eq:e2}
\end{equation}
results in the mechanical sub-model of DCV
\begin{eqnarray}
\label{eq:e3}
  \ddot{\nu} + 2\xi \omega_0 \, \dot{\nu} + \omega^2 \nu &=& \omega^2 u, \\
  z &=& h(\nu),
\label{eq:e4}
\end{eqnarray}
where the output $z$ is the orifice state which governs the valve
flow. The parameters $\alpha$ and $\beta$ denote the saturation
level and dead-zone size correspondingly. For the sake of
simplicity, we assume the same parameter values for both
directions of the spool displacement, hence a symmetric DCV, while
some more specific DCV assemblies can have different
direction-dependent $\alpha,\beta$ values, see e.g.
\cite{eryilmaz2006unified} for details. Further we note that while
the saturation level can be considered as constant and known
a-priory, or at least after system identification, the dead-zone
size can be subject to state-dependent uncertainties and altering
(wear-related) effects. The state-dependent variations lead back
to e.g. viscoelastic properties and clarity of the hydraulic
medium, working pressure, temperature, and others.

The volumetric flow of hydraulic medium associated with both
outlet ports of DCV, and thus with both chambers of the connected
hydraulic cylinder, is governed by the corresponding pressure
differences and given by the orifice equations
\begin{eqnarray}
\label{eq:e5}
  Q_A(z) & = & \left\{%
\begin{array}{ll}
    z K \sqrt{P_S - P_A}, & \hbox{ for } z > 0, \\[1mm]
    z K \sqrt{P_A - P_T}, & \hbox{ for } z < 0,\\[1mm]
    0,                    & \hbox{ otherwise;}
\end{array}%
\right.     \\[1mm]
    Q_B(z) & = & \left\{%
\begin{array}{ll}
    -z K \sqrt{P_B - P_T}, & \hbox{ for } z > 0, \\[1mm]
    -z K \sqrt{P_S - P_B}, & \hbox{ for } z < 0. \\[1mm]
    0,                    & \hbox{ otherwise.}
\end{array}%
\right. \label{eq:e6}
\end{eqnarray}
For the sake of clarity, we note that the configuration depicted
in Fig. \ref{fig:1} implies $z<0$, so that the (in) flow through
the port B has the positive sign, and the (out) flow through the
port A correspondingly negative. The valve flow coefficient
\begin{equation}
K = C_d w \sqrt{\frac{2}{\rho}}, \label{eq:e7}
\end{equation}
denoted also as a valve gain, depends of the oil density $\rho$,
discharge coefficient of the orifice $C_d$, and wight of
rectangular orifice area $w$. Last two are the structural
parameters of the DCV assembly at hand, cf. with e.g.
\cite{viall2000determining}. Also we note that while the flow
direction in (\ref{eq:e5}), (\ref{eq:e6}) is determined by the
sign of the orifice state $z$, including dead-zone, several works
decide the flow direction depending on the sign of control input
\cite{sohl1999experiments}, spool position
\cite{alleyne2000simplified,yao2000adaptive}, or pressure drop in
the chambers \cite{aranovskiy2013modeling}.

The continuity equations of pressure the gradient
\cite{merritt1967hydraulic} in both chambers are given by
\begin{eqnarray}
\label{eq:e8}
  \dot{P}_A &=& \frac{E}{V_A} \Bigl(Q_A - A_A \dot{x} - C_L(P_A-P_B)  \Bigr), \\
  \dot{P}_B &=& \frac{E}{V_B} \Bigl(Q_B + A_B \dot{x} - C_L(P_B-P_A)
  \Bigr),
\label{eq:e9}
\end{eqnarray}
where $A_{A/B}$ are the corresponding effective piston areas. Note
that for the single-rod cylinders $A_A > A_B$, cf. with Fig.
\ref{fig:1}, while for the double-rod cylinders $A_A = A_B$ is
mostly assumed. The bulk modulus $E= - V(dP/dV)$ reflects the
incompressibility of hydraulic medium in cylinder. While this is
generally pressure-dependent, an effective bulk modulus is mostly
assumed as a constant parameter for the rated operation pressure
of hydraulic system at hand. The total volume in hydraulic
circuits
\begin{eqnarray}
\label{eq:e10}
  V_A &=& V_A^0 + A_A \, \mathrm{sat}_{h}(x), \\
  V_B &=& V_B^0 - A_B \, \mathrm{sat}_{h}(x) \label{eq:e11}
\end{eqnarray}
depends each on the rod drive position, while $V_{A/B}^0$ is the
total chambers volume (including piping and fittings between the
DCV and cylinder) when the rod drive is in the initial zero
position $x = 0$. Note that the drive displacement is subject to
saturation, captured by $\mathrm{sat}_{h}$ operator in
(\ref{eq:e10}), (\ref{eq:e11}), while $h$ is the half-stroke of
the symmetric cylinder. $C_L$ is an internal leakage coefficient
of cylinder which characterizes an additional pressure drop due to
(minor) penetration of the hydraulic medium between both chambers.
That is caused by a non-zero clearance between the barrel and
piston of hydraulic cylinder. Note that in ideal case, the leakage
coefficient can be assumed as zero, though it constitutes a rather
uncertain factor related to the life cycle of hydraulic cylinder,
temperature-dependent material expansion, oil viscosity and, if
applicable, external radial stress affecting the piston rod during
the operation.

The mechanical sub-model of cylinder drive can be directly
formulated, for one translational degree-of-freedom (DOF), based
on the Pascal's and Newton's second law as
\begin{equation}
m\ddot{x} = P_A A_A - P_B A_B - f - F_{L}. \label{eq:e12}
\end{equation}
The lumped moving mass of the cylinder piston with rod and, if
applicable, external coupling (tool) is $m$, and all external
counteracting forces are summarized by $F_{L}$. Note that the
latter are the matter of application and, in the most simple case
of stand-alone hydraulic cylinder, can be assumed to be zero. The
counteracting nonlinear friction $f$ mainly depends on the
relative velocity of the piston drive and acts as an inherent
damping of the relative motion. Note that in more complex modeling
of the motion dynamics, the friction $f$ can be also considered as
depending on the normal load, relative displacement at motion
reversals, working temperature, and other factors. Furthermore,
the kinetic friction is well-known to exhibit a time-varying
behavior so that its model parameters can be assumed as drifting
or, at least, with an uncertainty range. For more details on
varying kinetic friction and its impact on the motion control we
refer to e.g. \cite{Ruderman15,ruderman2016integral}.

It can be stressed that the friction in hydraulic cylinders is
mainly due to the contacts between the rods and lip seals, and
between the piston o-rings, seals, and cylinder
\cite{owen2003reduction}. Also the viscous effects of hydraulic
fluid contribute to arising friction which counteracts the induced
motion of the cylinder drive. However, the major source of
friction remains due to the tight lip and piston seals that are
required to prevent internal and external leaks in the system. In
the following, we will assume the kinetic friction as function of
solely the relative velocity $\dot{x}$, while some works
\cite{bonchis1999pressure,ottestad2012reducing} made attempts to
explicitly incorporate dependency on the pressure, correspondingly
pressure difference, into the friction law. Further we note that
more sophisticated dynamic friction models can also capture some
transient and memory-driven effects of nonlinear friction, see
\cite{armstrong1994survey} for survey, and pre-sliding
rate-independent damping \cite{RudRach2017} as well. The
steady-state friction can be described by the well-known Stribeck
\cite{Stribeck1902} characteristic curve
\begin{equation}
f(\dot{x})=\mathrm{sign}(\dot{x})\Bigl(F_{c}+(F_{s}-F_{c})
\exp\bigl(-|\dot{x}|^{\delta} \chi^{-\delta} \bigr)\Bigr) + \sigma
\dot{x}, \label{eq:e13}
\end{equation}
which takes into account the constant Coulomb and linear viscous
friction, plus the velocity-weakening effects in a low velocity
range around zero. The static model (\ref{eq:e13}) is
parameterized by the Coulomb friction coefficient $F_{c}>0$,
Stribeck (or stiction) friction coefficient $F_{s} > F_{c}$,
linear viscous friction coefficient $\sigma > 0$, and two Stribeck
shape factors $\chi > 0$ and $\delta \neq 0$. In case the
discontinuity at velocity zero crossing should be avoided in the
model, to say without performing a more complex dynamic friction
modeling, some heuristic approaches e.g. using tangens-hyperbolic
functions instead of the sign-operator can be pursued, see for
example \cite{yao2015adaptive}. Another reason for using the
tangens-hyperbolic type smooth transitions, instead of the $sign$
discontinuity, is that during fast transients the argument of
square root in (\ref{eq:e5}), (\ref{eq:e6}) can yield temporary
negative, thus making calculus improper. Several characteristic
shapes of the steady-state friction determined in experiments on
the hydraulic cylinder drives can be found in
\cite{tafazoli1995friction,lischinsky1999friction,alleyne2000simplified,owen2003reduction,marton2011practical,ottestad2012reducing}.
It should be noted that the above friction parameters can be also
subject to uncertainties dictated by the thermal and load
conditions, life cycle and wear, dust, dwell-time, spatial
properties of contacting surfaces and others.

\section{REDUCED MODEL}
\label{sec:3}

The reduced model takes advantages of the fast DCV spool dynamics
so that the second-order closed-loop behavior of DCV can be
neglected. Alternatively, this can be approximated by some
constant time delay, if the corresponding phase lag is still to be
taken into account when designing a feedback control of the
hydraulic cylinder. Further, the cylinder piston velocity is
assumed as the system output of interest, since this can be either
directly derived from the available position measurement or
observed, correspondingly estimated, using various robust velocity
estimation techniques known from the literature, e.g.
\cite{levant1998robust,davila2005second}. In the following, we
assume that the time-constants of hydraulic and mechanical
sub-dynamics are significantly larger than that of the DCV control
loop, to say by two to three orders of magnitude. Consequently,
the input of the reduced model can be assumed as $u^{*} = v$ which
directly enters both nonlinearities according to (\ref{eq:e2}).
Therefore, the resulted reduced model is with static
nonlinearities in the input channel.

Introducing the load-related pressure $P_L = P_A-P_B$ and assuming
a closed hydraulic circuit, i.e. $|Q_A|=|Q_B|$, the orifice
equations (\ref{eq:e5}), (\ref{eq:e6}) can be aggregated into one
\begin{equation}
Q_L = z K \sqrt{\frac{1}{2}\Bigl( P_S - \mathrm{sign}(z) P_L
\Bigr)}, \label{eq:e14}
\end{equation}
while it is valid
\begin{equation}
P_A = \frac{P_S+P_L}{2}, \quad  P_B = \frac{P_S-P_L}{2}.
\label{eq:e15}
\end{equation}
Note that (\ref{eq:e14}) represents an average load flow through
the DCV and occurs as further nonlinearity in the input channel of
the reduced model. This incorporates, however, a dynamic feedback
of the load pressure $P_L$ so that only (\ref{eq:e2}) can be
considered as a static input nonlinearity in forward. Furthermore
it should be stressed that (\ref{eq:e14}), (\ref{eq:e15}) assume
also zero pressure in the tank, cf. with (\ref{eq:e5}),
(\ref{eq:e6}), which is however a reasonable simplification for
various hydraulic cylinder drive systems.

Following the above way of aggregation, the hydraulic continuity
equations (\ref{eq:e8}), (\ref{eq:e9}) can be transformed into one
related to the load pressure gradient
\begin{equation}
\dot{P}_L = \frac{4E}{V_{t}} \Bigl(Q_L - \bar{A} \dot{x} - C_L P_L
\Bigr). \label{eq:e16}
\end{equation}
Here the total hydraulic actuator volume is $V_t = V_A^0 + V_B^0$
and the average effective piston area is $\bar{A} = 0.5
(A_A+A_B)$. Note that the latter yields an exact value for
double-rod cylinders, while for single rod cylinders it bears an
averaging error of the half of the rod cross-section area.

Correspondingly, the dynamics (\ref{eq:e12}) of mechanical part
transforms (for the reduced model) into
\begin{equation}
m \frac{d}{dt} \dot{x} = P_L \bar{A} - f(\dot{x}) - F_{L},
\label{eq:e17}
\end{equation}
with the cylinder drive velocity $\dot{x}$ as system output and
total friction force given by (\ref{eq:e13}). Here we recall that
the reduced model (\ref{eq:e13})-(\ref{eq:e17}) with the input
nonlinearity (\ref{eq:e2}), (\ref{eq:e4}) has the second-order
dynamics, while the external load force $F_L$ acts as an internal
state disturbance interfering between the first and second
integrators of the forward dynamics path.

\section{ANALYSIS OF SYSTEM DYNAMICS}
\label{sec:4}

In the following, we discuss dynamic properties of both models
while making some qualitative comparisons between them and
demonstrating the differences in terms of the measured frequency
response functions and state trajectories. For the latter we
consider the $[P_L,\, \dot{x}]$ space, since both dynamic states
are the most significant and characteristic for the response of
hydraulic, correspondingly mechanic, parts of cylinder drives
under an actuation. Before comparing the models, we provide a
detailed analysis of the linearized reduced-order dynamics so as
to disclose the principal system behavior with impact of
linearization.
\begin{table}[!h]
  \renewcommand{\arraystretch}{1.1}
  \caption{System parameters of numerical simulation}
  \footnotesize
  \label{tab:1}
  \begin{center}
  \begin{tabular} {||p{1cm}|p{1cm}|p{1cm}||p{1cm}|p{1cm}|p{1cm}||}
  \hline
  Param.     &   Value   &  Units    &   Param.  &  Value    &  Units \\
  \hline \hline
  $\alpha$   &   1e-3    &  m        & $A_A$     &   5e-3    &   m$^2$ \\
  \hline
  $\beta$    &   3e-4    &  m        & $A_B$     &   4.7e-3  &   m$^2$ \\
  \hline
  $\omega$   &   1200    &  rad/s    & $V_A^0$   &   1.2e-3  &   m$^3$ \\
  \hline
  $\xi$      &   0.7     &  none     & $V_B^0$   &   1.15e-3 &   m$^3$ \\
  \hline
  $C_d$      &   0.65    &  none     & $m$       &   20      &   kg \\
  \hline
  $w$        &   0.02    &  m        & $F_c$     &   600     &   N \\
  \hline
  $\rho$     &   850     &  kg/m$^3$ & $F_s$     &   900     &   N \\
  \hline
  $E$        &   1e8     &  Pa       & $\sigma$  &   2000    &   kg/s \\
  \hline
  $P_S$      &   1e7     &  Pa       & $\chi$    &   0.02    &   m/s \\
  \hline
  $P_T$      &   0       &  Pa       & $\delta$  &   0.8     &  none \\
  \hline
  $C_L$      &   0       &  1/s      & $h$       &   0.2     &  m  \\
  \hline
  \end{tabular}
  \end{center}
  \normalsize
\end{table}
For the accompanying numerical simulations, the system parameters
listed in Table \ref{tab:1} are assumed. All units are given in
SI. The numerical values are assigned artificially but, at the
same time, lie in the range of realistic values for the DVC
controlled hydraulic cylinder actuators, cf. with parameters
provided e.g. in \cite{owen2003reduction,marton2011practical}.

\subsection{Linearized reduced-order dynamics}
\label{sec:4:sub:1}

The purpose of a DCV is to supply hydraulic cylinders with the
controlled volumetric flow which energizes the hydraulic circuits
and mechanical drive according to (\ref{eq:e8}), (\ref{eq:e9}),
(\ref{eq:e12}), correspondingly (\ref{eq:e16}), (\ref{eq:e17}).
Therefore, to obtain first an insight into dynamic behavior of the
system, i.e. in frequency domain, the reduced-order model can be
directly linearized between the load flow and rod velocity as
\begin{equation}
G(s) = \frac{\dot{x}(s)}{Q_L(s)} = \frac{ \bar{A}^{-1}
}{\omega_c^{-2} \, s^2 + 2 \zeta \omega_c^{-1} \, s +1},
\label{eq:e18}
\end{equation}
with the cylinder eigenfrequency and damping given by
\begin{eqnarray}
\label{eq:e19}
  \omega_c &=& 2 \bar{A} \sqrt{\frac{E}{V_t m}}, \\[1mm]
  \zeta    &=& \frac{\sigma}{4\bar{A}} \sqrt{\frac{V_t}{E m}}.
\label{eq:e20}
\end{eqnarray}
Note that the single step required to obtain the linearized
dynamics (\ref{eq:e18}) is that of assuming zero leakage
coefficient and neglecting the nonlinear part of the system
friction. The latter implies that a relative motion of cylinder
rod becomes solely damped by the viscous friction with a linear
damping coefficient $\sigma$. This represents an artificial case
of the weakest mechanical damping, while a real dynamics of
hydraulic cylinder is subject to additional rate-independent
damping due to the Coulomb and presliding friction,
correspondingly stiction.

Despite the above simplification, the linearized dynamics
(\ref{eq:e18})-(\ref{eq:e20}) offers a creditable estimation of
the second-order system behavior. This is, above all, in terms of
the system gain, eigenfrequency and therefore system stiffness,
and damping which reveals the system as stronger or less
oscillatory and has a direct impact on the closed-loop behavior
when designing a feedback control. It is evident that the gain
between the flow and cylinder velocity is inverse to the effective
piston area, so that the cylinders with larger $\bar{A}$ can
achieve higher velocities. Also the eigenfrequency and therefore
achievable control bandwidth increase with $\bar{A}$, according to
(\ref{eq:e19}). At the same time, one should keep in mind that
increasing $\bar{A}$ proportionally reduces the system damping,
see (\ref{eq:e20}), and therefore evokes additional issues of
system stability. Furthermore from (\ref{eq:e19}), one can see
that both the moving mass and total hydraulic volume constitute
the inertial terms which slow down the system eigenfrequency. From
the energy transfer point of view it appears as logically
consistent, since a higher hydraulic volume is more inertial,
correspondingly with higher amount of kinetic energy for the same
displacement rate. The bulk modulus $E$ appears as an equivalent
stiffness of hydraulic medium so that larger $E$ increases the
system eigenfrequency, according to (\ref{eq:e19}). However,
unlike the mechanical stiffness, the bulk modulus influences the
system damping as well, cf. with (\ref{eq:e20}). One can sum up
that the parameters pair $E/V_t$, in addition to the effective
piston area, is decisive for the system eigenfrequency and
damping, as related to hydraulic circuits. This is crucial for
specifying and dimensioning of hydraulic cylinder actuators during
the design. Also it is obvious that the friction is the single
factor influencing the overall system damping without having
inverse impact on its eigenfrequency.

The associated closed-loop dynamics for system (\ref{eq:e18}) can
be analyzed when examining the root locus of the open-loop $k
G(s)s^{-1}$. Here the position of cylinder rod is taken as output
of interest (used for a feedback control), and $k$ is a system
feedback gain.
\begin{figure}[!h]
\centering
\includegraphics[width=0.85\columnwidth]{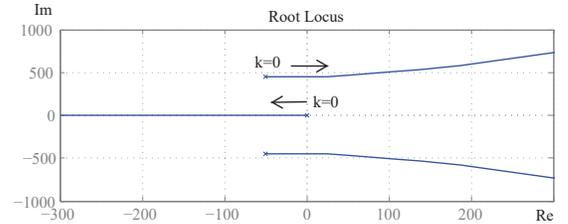}
\caption{Root locus for the open-loop $k G(s)s^{-1}$.}
\label{fig:2}
\end{figure}
The root locus diagram, depicted in Fig. \ref{fig:2} for the
parameters from Table \ref{tab:1}, discloses that the system
remains oscillating due to the conjugate-complex pole pair and
become unstable starting from certain feedback gain value.

\begin{figure}[!h]
\centering
\includegraphics[width=0.85\columnwidth]{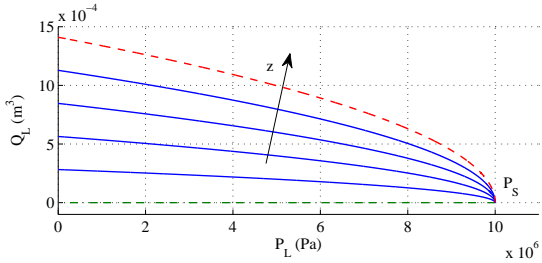}
\caption{Static load flow-pressure characteristic curves.}
\label{fig:3}
\end{figure}

Next the attention is to be paid to the impact of the load
pressure feedback on the load flow according to (\ref{eq:e14}).
Keeping constant $z$ value as a DCV operation point, the
flow-pressure characteristics can be directly computed while $\max
P_L = P_S$. The static flow/pressure characteristic curves for
different $z$-values, starting from the neutral (closed) orifice
state $z=0$ and going until its maximal value $z=\alpha$, are
depicted in Fig. \ref{fig:3} for the assumed system parameters
from Table \ref{tab:1}. Note that for negative load values the
depicted characteristic curves are symmetrical with respect to
both axes and lie in the third quadrant for the negative
$(P_L,\,Q_L)$ pairs. Considering the working points $\hat{z}$ and
$\hat{P}_{L}$ and positive DCV operation range, this without loss
of generality, one can show that
\begin{eqnarray} \label{eq:e21}
  \frac{\partial Q_L}{\partial z}  \Bigr|_{\hat{P}_L}     &=& K \sqrt{P_S - \hat{P}_L} =: C_q, \\[1mm]
  \frac{\partial Q_L}{\partial P_L} \Bigr|_{\hat{z}}   &=& - \frac{K \hat{z}}{2 \sqrt{P_S - \hat{P}_L}} =: -C_{qp}.
\label{eq:e22}
\end{eqnarray}
Both partial derivatives allow to introduce the so called flowgain
coefficient $C_q$ and flow-pressure coefficient $C_{qp}$. Then,
for linearizing the orifice equation (\ref{eq:e14}), one can write
\begin{equation}
\hat{Q}_L = C_q z - C_{qp} P_L. \label{eq:e23}
\end{equation}
It is evident that the pressure-dependent coefficient $C_q$
amplifies the input orifice state $z$ and, therefore, introduces
solely an additional gain to the transfer function (\ref{eq:e18}).
On the contrary, the pressure feedback coefficient $C_{qp}$
reshapes the $G(s)$ dynamics so that the frequency response
characteristics changes depending on the operation point. Already
from the characteristic curves shown in Fig. \ref{fig:3}, one can
see that the feedback gain $C_{qp}$, which is a slope of the
tangent to the $(Q_L,P_L)$ curve for each $(\hat{z},\hat{P}_L)$
working point, is rather low for small amplitudes. For zero load
flows, i.e. at neutral DCV state with zero orifice, the feedback
gain remains zero even for higher load pressures. This explain why
the hydraulic cylinders under pressure often exhibit a
steady-state vibration noise, even when no apparent motion of
cylinder, i.e. no load flow, occur.

\begin{figure}[!h]
\centering
\includegraphics[width=0.98\columnwidth]{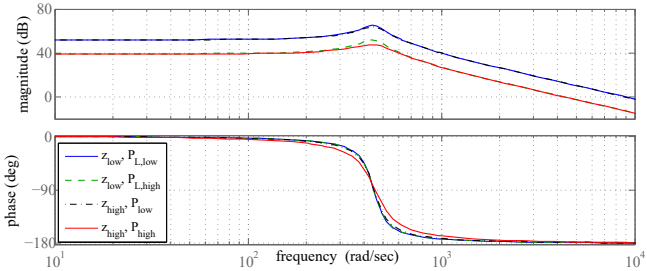}
\caption{FRFs of linearized system (\ref{eq:e24}) at different
$(\hat{z},\, \hat{P}_L)$ operation points.} \label{fig:4}
\end{figure}
For better exposition of the above analysis consider the
linearized transfer function between the orifice state $z$ and rod
velocity $\dot{x}$, which can be obtained from
(\ref{eq:e18})-(\ref{eq:e20}) and (\ref{eq:e23}) as
\begin{equation}
\hat{G}(s) = \frac{\dot{x}(s)}{z(s)} = \frac{C_q
G(s)}{1+C_{qp}(ms+\sigma)\bar{A}^{-1}G(s)}. \label{eq:e24}
\end{equation}
Obviously, for zero flow-pressure coefficient, i.e. for no load
pressure feedback, the (\ref{eq:e24}) transforms to the previously
considered dynamics (\ref{eq:e18}) amplified by the flow gain
$C_q$. To provide a quantitative comparison between the linearized
dynamics (\ref{eq:e24}) at different operation points assume four
pairs by combining the relatively low and high values
$\hat{z}_{low,high}=\{0.05\alpha, \, 0.95 \alpha \}$ and
$\hat{P}_{L,low,high}=\{0.05 P_S, \, 0.95 P_S \}$. Recall that
$\alpha$ and $P_S$ constitute the upper bounds for $z$ and $P_L$
correspondingly. The frequency response functions (FRFs) for all
four $(\hat{z},\, \hat{P}_L)$ combinations are shown opposite to
each other in Fig. \ref{fig:4}. One can see that the low and high
load pressure values change significantly the system gain, while
the principal shape of the frequency response, with corresponding
eigenfrequency and damping, remains quite similar. It can be noted
that an increased system damping is solely in case of the maximal
orifice and load pressure (red solide line).

\subsection{Magnitude-dependent frequency response}
\label{sec:4:sub:2}

In what follows, the numerical simulations of the full-order and
reduced models, according to Sections \ref{sec:2}, \ref{sec:3},
have been used while assuming the parameters from Table
\ref{tab:1} and Forward-Euler integration with a fixed step
(0.0001) solver of MathWorks Simulink$^{\circledR}$ software. The
$\mathrm{sign}(\dot{x})$ in (\ref{eq:e13}) has been replaced by
$\tanh(400\dot{x})$ so as to capture the negative pressure
differences under square root in the orifice equations.

\begin{figure}[!h]
\centering
\includegraphics[width=0.98\columnwidth]{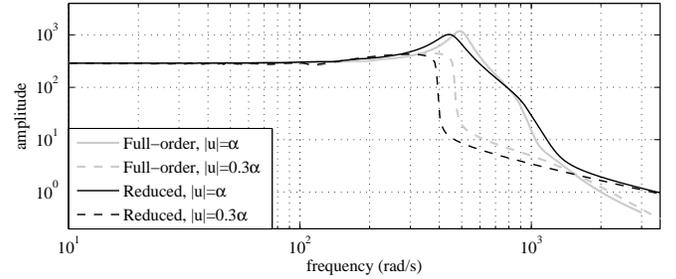}
\caption{FRFs of full-order and reduced models at different input
amplitudes.} \label{fig:5}
\end{figure}
The FRFs of the full-order and reduced models have been obtained
from the numerical simulation at different excitation amplitudes,
$|u| = \{\alpha, \, 0.3\alpha \}$, when applying a down-chirp
600--1 Hz and excluding the input nonlinearities (\ref{eq:e2}).
Estimated with a standard H1 correlation algorithm and smoothed,
the FRFs are depicted opposite to each other in Fig. \ref{fig:5}.
One can see that for both models the low inputs do not excite the
hydraulic-related resonance peak. The resonance peak of the
reduced model is also shifted to the lower frequencies. At higher
frequencies the reduced model has inherently lower, $-40$ dB per
decade, decrease due to the second order dynamics.

\subsection{State trajectories}
\label{sec:4:sub:3}

\begin{figure}[!h]
\centering
\includegraphics[width=0.48\columnwidth]{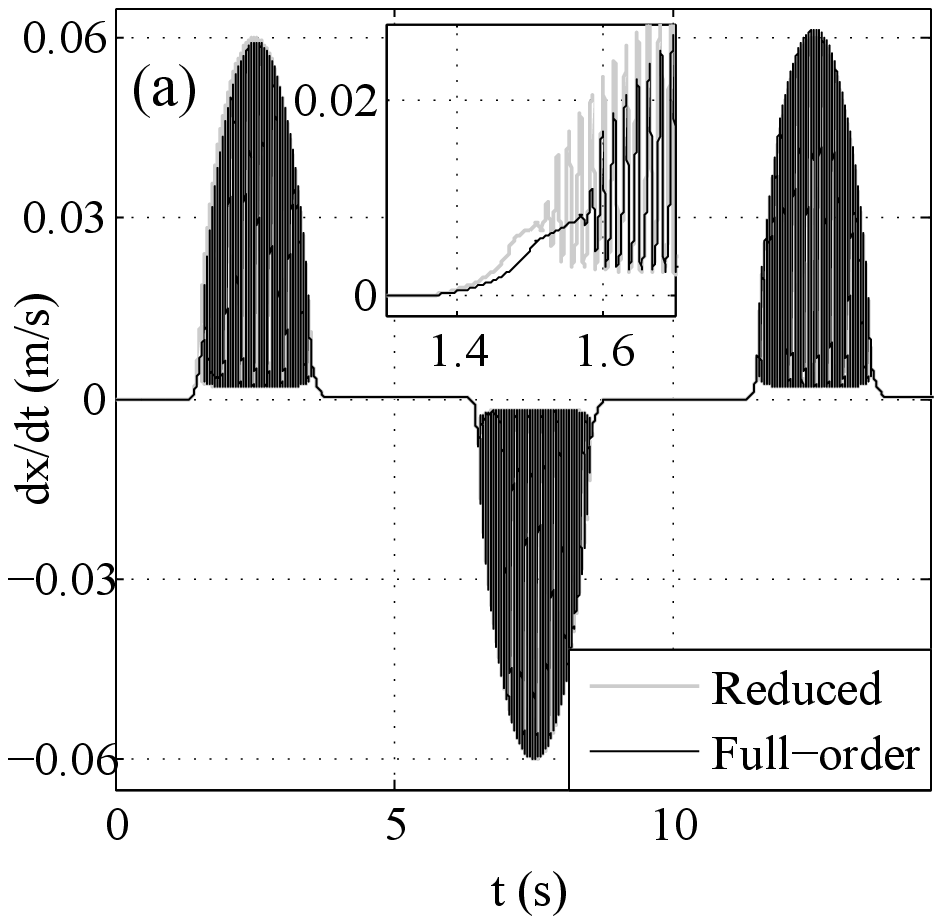}
\includegraphics[width=0.48\columnwidth]{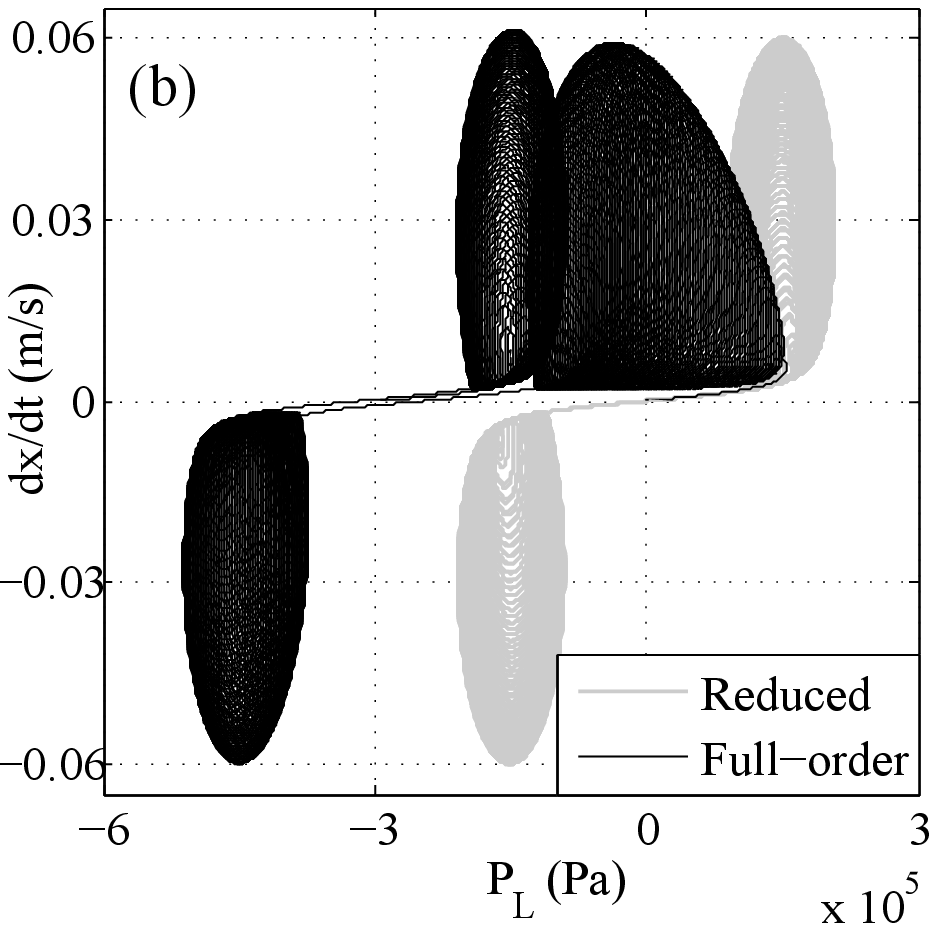}
\includegraphics[width=0.98\columnwidth]{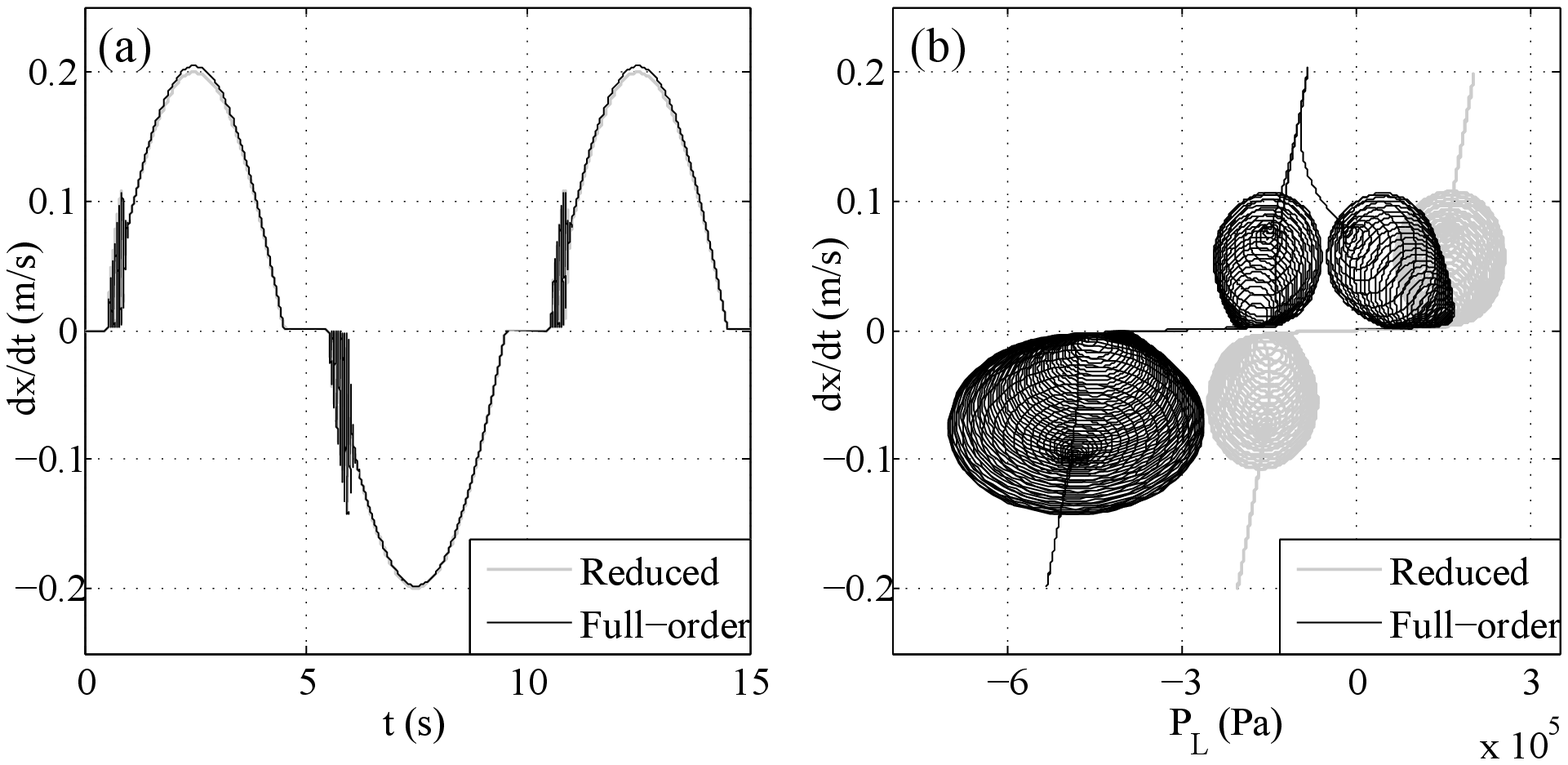}
\caption{Velocity trajectories (a) and phase plane portraits (b)
for the sinusoidal inputs with low (upper) and high (below)
amplitudes.} \label{fig:6}
\end{figure}

The $(P_L, \dot{x})$ state trajectories are analyzed, for both
models, when applying a low sinusoidal amplitude 0.4e-3, which is
close to the dead-zone size $\beta$, and high sinusoidal amplitude
1e-3, which corresponds to the saturated orifice state. Both input
sinusoidals are at 0.1 Hz frequency without phase shift. The
velocity responses and phase plane portraits of both models are
shown in Fig. \ref{fig:6}, for the low (on the top) and high (on
the bottom) excitation amplitudes. One can see that while the
velocity patterns of both models coincide well with each other,
the load pressure trajectories are quite different in both,
transient oscillations and steady-state locations. This is not
surprising since at relatively low load pressures, the reduced
model averaging of the orifice and continuity equations yields the
hydraulic circuits as ideally symmetrical.
On the contrary, the full-order model takes into account the
one-side rod cylinder that reflects in the shifted (unbalanced)
load pressure during a cyclic open-loop excitation. From both,
velocity trajectories and phase plane portraits, one can see zero
velocity regions corresponding to the DCV dead-zone. Further one
can recognize the transient oscillations of both states in the
anti-phase each time the relative motion restarts in an opposite
direction. After the fast transients, the $(P_L,
\dot{x})$-trajectory attains a non-oscillating pattern which
corresponds to the steady-state motion in case of the higher
excitation amplitude.

\section{CONCLUSIONS}
\label{sec:5}

The full- and reduced-order dynamic models for DCV controlled
hydraulic cylinders have been described in details. A possible
reduction from the 5th to the 2nd order dynamics has been provided
while explaining the related assumptions and implications. The
system behavior has been analyzed and the basic equations have
been exposed in terms of their parametrization and coupling
between each other. Numerical examples with parameters assumed
close to the real hydraulic servo systems have been considered,
with particular focus on the open-loop behavior in frequency
domain and trajectories of the most relevant system states, load
pressure and cylinder velocity. The given modeling and analysis
should serve for better understanding the dynamics of hydraulic
cylinder drives and associated identification and motion control
design.

\section*{Acknowledgment}
This work has received funding from the European Union Horizon
2020 research and innovation programme H2020-MSCA-RISE-2016 under
the grant agreement No 734832.

\bibliographystyle{IEEEtran}
\bibliography{references}

\begin{thebibliography}{10}
\providecommand{\url}[1]{#1}
\csname url@rmstyle\endcsname
\providecommand{\newblock}{\relax}
\providecommand{\bibinfo}[2]{#2}
\providecommand\BIBentrySTDinterwordspacing{\spaceskip=0pt\relax}
\providecommand\BIBentryALTinterwordstretchfactor{4}
\providecommand\BIBentryALTinterwordspacing{\spaceskip=\fontdimen2\font plus
\BIBentryALTinterwordstretchfactor\fontdimen3\font minus
  \fontdimen4\font\relax}
\providecommand\BIBforeignlanguage[2]{{%
\expandafter\ifx\csname l@#1\endcsname\relax
\typeout{** WARNING: IEEEtran.bst: No hyphenation pattern has been}%
\typeout{** loaded for the language `#1'. Using the pattern for}%
\typeout{** the default language instead.}%
\else
\language=\csname l@#1\endcsname
\fi
#2}}

\bibitem{merritt1967hydraulic}
H.~E. Merritt, \emph{Hydraulic control systems}.\hskip 1em plus 0.5em minus
  0.4em\relax John Wiley \& Sons, 1967.

\bibitem{yao2000adaptive}
B.~Yao, F.~Bu, J.~Reedy, and G.-C. Chiu, ``Adaptive robust motion control of
  single-rod hydraulic actuators: theory and experiments,'' \emph{IEEE/ASME
  Trans. on Mechatronics}, vol.~5, no.~1, pp. 79--91, 2000.

\bibitem{alleyne2000simplified}
A.~Alleyne and R.~Liu, ``A simplified approach to force control for
  electro-hydraulic systems,'' \emph{Control Engineering Practice}, vol.~8,
  no.~12, pp. 1347--1356, 2000.

\bibitem{komsta2013integral}
J.~Komsta, N.~van Oijen, and P.~Antoszkiewicz, ``Integral sliding mode
  compensator for load pressure control of die-cushion cylinder drive,''
  \emph{Control Engineering Practice}, vol.~21, no.~5, pp. 708--718, 2013.

\bibitem{sohl1999experiments}
G.~A. Sohl and J.~E. Bobrow, ``Experiments and simulations on the nonlinear
  control of a hydraulic servosystem,'' \emph{IEEE Trans. on Control Systems
  Technology}, vol.~7, no.~2, pp. 238--247, 1999.

\bibitem{marton2011practical}
L.~M{\'a}rton, S.~Fodor, and N.~Sepehri, ``A practical method for friction
  identification in hydraulic actuators,'' \emph{Mechatronics}, vol.~21, no.~1,
  pp. 350--356, 2011.

\bibitem{aranovskiy2013modeling}
S.~Aranovskiy, ``Modeling and identification of spool dynamics in an industrial
  electro-hydraulic valve,'' in \emph{IEEE 21st Mediterranean Conference on
  Control \& Automation (MED)}, 2013, pp. 82--87.

\bibitem{eryilmaz2006unified}
B.~Eryilmaz and B.~H. Wilson, ``Unified modeling and analysis of a proportional
  valve,'' \emph{Journal of the Franklin Institute}, vol. 343, no.~1, pp.
  48--68, 2006.

\bibitem{viall2000determining}
E.~N. Viall and Q.~Zhang, ``Determining the discharge coefficient of a spool
  valve,'' in \emph{American Control Conference (ACC2000)}, vol.~5, 2000, pp.
  3600--3604.

\bibitem{Ruderman15}
M.~Ruderman and M.~Iwasaki, ``Observer of nonlinear friction dynamics for
  motion control,'' \emph{IEEE Trans. on Industrial Electronics}, vol.~62,
  no.~9, pp. 5941--5949, 2015.

\bibitem{ruderman2016integral}
M.~Ruderman, ``Integral control action in precise positioning systems with
  friction,'' in \emph{IFAC 12th Workshop on Adaptation and Learning in Control
  and Signal Processing}, 2016, pp. 82--86.

\bibitem{owen2003reduction}
W.~S. Owen and E.~A. Croft, ``The reduction of stick-slip friction in hydraulic
  actuators,'' \emph{IEEE/ASME Trans. on Mechatronics}, vol.~8, no.~3, pp.
  362--371, 2003.

\bibitem{bonchis1999pressure}
A.~Bonchis, P.~I. Corke, and D.~C. Rye, ``A pressure-based, velocity
  independent, friction model for asymmetric hydraulic cylinders,'' in
  \emph{IEEE International Conference on Robotics and Automation}, vol.~3,
  1999, pp. 1746--1751.

\bibitem{ottestad2012reducing}
M.~Ottestad, N.~Nilsen, and M.~R. Hansen, ``Reducing the static friction in
  hydraulic cylinders by maintaining relative velocity between piston and
  cylinder,'' in \emph{12th International Conference on Control, Automation and
  Systems (ICCAS)}, 2012, pp. 764--769.

\bibitem{armstrong1994survey}
B.~Armstrong-H{\'e}louvry, P.~Dupont, and C.~C. De~Wit, ``A survey of models,
  analysis tools and compensation methods for the control of machines with
  friction,'' \emph{Automatica}, vol. 30(7), pp. 1083--1138, 1994.

\bibitem{RudRach2017}
M.~Ruderman and D.~Rachinskii, ``Use of {Prandtl-Ishlinskii} hysteresis
  operators for {Coulomb} friction modeling with presliding,'' in \emph{Journal
  of Physics: Conference Series}, vol. 811, no.~1, 2017, p. 012013.

\bibitem{Stribeck1902}
R.~Stribeck, ``Die wesentlichen {Eigenschaften} der {Gleit-} und
  {Rollenlager},'' \emph{VDI-Zeitschrift (in German)}, vol.~46, no. 36--38, pp.
  1341--1348,1432--1438,1463--1470, 1902.

\bibitem{yao2015adaptive}
J.~Yao, W.~Deng, and Z.~Jiao, ``Adaptive control of hydraulic actuators with
  {LuGre} model-based friction compensation,'' \emph{IEEE Tran. on Industrial
  Electronics}, vol.~62, no.~10, pp. 6469--6477, 2015.

\bibitem{tafazoli1995friction}
S.~Tafazoli, C.~De~Silva, and P.~Lawrence, ``Friction estimation in a planar
  electrohydraulic manipulator,'' in \emph{American Control Conference
  (ACC'95)}, vol.~5, 1995, pp. 3294--3298.

\bibitem{lischinsky1999friction}
P.~Lischinsky, C.~Canudas-de Wit, and G.~Morel, ``Friction compensation for an
  industrial hydraulic robot,'' \emph{IEEE Control Systems Magazine}, vol.~19,
  no.~1, pp. 25--32, 1999.

\bibitem{levant1998robust}
A.~Levant, ``Robust exact differentiation via sliding mode technique,''
  \emph{Automatica}, vol.~34, no.~3, pp. 379--384, 1998.

\bibitem{davila2005second}
J.~Davila, L.~Fridman, and A.~Levant, ``Second-order sliding-mode observer for
  mechanical systems,'' \emph{IEEE Trans. on Automatic Control}, vol.~50,
  no.~11, pp. 1785--1789, 2005.

\end{thebibliography}

\end{document}